\documentclass[aip,pop,reprint,amsmath,amssymb,nofootinbib]{revtex4-1}
\usepackage{geometry} %%MESSES UP THE REVTEX FORMAT'S SPACING
\usepackage{color}
\usepackage{calc}
\usepackage{amsmath}
\usepackage{amssymb}
\usepackage{cancel}
\usepackage{ulem}
\usepackage{sidecap}
\makeatletter
\usepackage{bm}
\usepackage{soul}% for highlighting
\usepackage{url}
\usepackage{varioref}
\usepackage{graphics}
\usepackage{subfigure}
\usepackage{adjustbox}
\usepackage{appendix}
\usepackage{multirow}
\usepackage{verbatim}
\usepackage{colortbl}
\usepackage{times}
\usepackage{multimedia}
\usepackage{enumerate}
\usepackage{dcolumn}
\usepackage[usenames,dvipsnames,svgnames,table]{xcolor}
\usepackage{wrapfig}
\usepackage{nth}
\usepackage{textgreek,upgreek}
\newcommand{\f}{\frac}

\newcommand{\mf}{\mathbf}
\newcommand{\be}{\begin{equation}}
\newcommand{\ba}{\begin{eqnarrray}}
\newcommand{\ee}{\end{equation}}
\newcommand{\ea}{\end{eqnarray}}

\def\th{\theta}

\def\part{\partial}

\newcommand{\bmin}{\begin{minipage}{0.495\textwidth}}
\newcommand{\emin}{\end{minipage}}

\def\Talf{\tau_{A}}

\newcommand{\vvec}{\mathbf{v}}
\newcommand{\Bvec}{\mathbf{B}}
\newcommand{\Evec}{\mathbf{E}}

\newcommand{\Jvec}{\mathbf{J}}

\def\vpar{v_{\parallel}}

\def\psiw{\tilde{\psi}_w}
\def\psit{\tilde{\psi}_t}
\def\tpsi{\tilde{\psi}}
\def\psic{\tilde{\psi}_{c}}

\def\divV{\nabla\cdot\vvec}
\def\capom{\Omega}

\begin{document}
\title{Nonlinear error field response in the presence of plasma rotation and real frequencies due to favorable curvature}
\author{Cihan Ak\c{c}ay}\email{c\_akcay@tibbartech.com}
\author{John M.~Finn} 
\affiliation{Tibbar Plasma Technologies, 274 DP Rd., Los Alamos, NM 87544}
\author{Andrew J.~ Cole}
\affiliation{Columbia University}
\author{Dylan P.~Brennan}
\affiliation{Princeton University}
%\maketitle
%\tableofcontents

\begin{abstract}
We present nonlinear resistive MHD simulations of the response of a rotating plasma to an error field when the plasma has weakly damped linear tearing modes (TM's), stabilized by pressure gradient and favorable curvature. Favorable curvature leads to the Glasser effect: the occurrence of real frequencies and stabilization with positive stability index $\Delta'$. A cylinder with hollow pressure is used to model the toroidal favorable curvature. Linear simulations with rotation and an error field $\psiw$ show, in agreement with analytic results, that the peak reconnected flux occurs for a rotation rate near the TM phase velocity. Nonlinear simulations with small $\psiw$ show that the real frequency and stabilization by favorable average curvature are masked by a nonlinear effect that occurs for very thin islands: flattening of the pressure across the island, mainly due to sound wave propagation. This flattening causes the disappearance of real frequencies destabilization of the mode, allowing it to grow to large amplitude similar to a β=0 unstable TM. The flattening of the current for larger islands saturates the mode nonlinearly. In the post-saturation phase, the interaction of the error field with the destabilized spontaneous tearing mode, which rotates with the plasma, leads to oscillations in the Maxwell torque and therefore modulations in the plasma rotation. The islands also rotate with  modulated phase velocity, undergoing small-amplitude oscillations due to these modulations. We also present a quasilinear model with an unstable spontaneous TM and error fields, showing that the superposition of these fields results in similar oscillations.

\end{abstract}
\maketitle % For REVTEX4.1

\section{Introduction\label{sec:Introduction}}
The MHD response of a rotating toroidal plasma to non-axisymmetric error fields\cite{Fitzy1991,Fitzy1993}, a form of driven magnetic reconnection, can be important in tokamaks because it can cause disruptions. 
The so-called error fields at the plasma edge can arise due to imperfections or alignment errors in external coils or disadvantageously placed current feeds. 
Such non-axisymmetric fields can also be applied at the boundary for various purposes, for example to provide resonant magnetic perturbations (RMPs) for mitigating edge-localized modes (ELM) \cite{Liu2012}. These fields exert a Maxwell torque on a rotating plasma and can lead to locking of the plasma rotation \cite{Fitzy1993}, a major cause of disruptions.
Therefore, for advancing tokamak science it is crucial  to understand the response of a plasma to non-axisymmetric external magnetic fields in the presence of plasma rotation. 

In driven magnetic reconnection, the response to error fields is largest when the associated tearing mode is weakly stable.  
This response has been observed \cite{Finn2015} to be qualitatively different for tearing layers with real frequencies in the plasma frame $\pm \omega_r$ from those with $\omega_r=0$. 
Real frequencies $\pm\omega_{r}$ can occur in resistive MHD tearing layers in the presence of favorable average curvature, a toroidal effect, in the resistive-inertial (RI) tearing regime.
This effect is called the Glasser or Glasser-Greene-Johnson (GGJ) effect \cite{Glasser1975,Glasser1976}, which shows that the real frequencies occur for $0<\Delta_c<\Delta'<\Delta_{m}$, and stabilization of the mode occurs for for $0<\Delta'<\Delta_c<\Delta_m$, where $\Delta_c$ and $\Delta_m$ are parameters related to average curvature. 
It has recently been shown in Ref.~\onlinecite{Finn2017} that a similar behavior occurs in the visco-resistive (VR) resistive MHD regime as well as in the RI regime, suggesting that the mode propagation and stabilization
occur over a wider range of parameters than previously realized. This effect in either of these two resistive MHD regimes is much like the behavior of the two-fluid drift-tearing mode \cite{Coppi1965,Biskamp1978,Bussac1978,FinnManheimerAntonsen}, which introduces diamagnetic propagation $\omega_r\sim\omega_*$ and stabilization. 
A major distinction is that in the RI and VR regimes, the modes occur in conjugate pairs, with $\pm \omega_r$, whereas in two-fluid theory, there is a single mode with $\omega_r \sim \omega_*$.

%Recent results \cite{Finn2015} focused on tearing layers with $\omega_r \neq 0$ and with $E\times B$ plasma rotation. 
In the presence of an error field represented by the flux $\psiw$, the response, i.e.~the reconnected flux $\psit$ at the tearing layer, was observed to be largest when the real frequency of the mode is Doppler-shifted by  $E\times B$ to a frequency that is close to zero in the lab frame: $\pm \omega_r +\mf{k}\cdot\vvec=\pm \omega_r+\Omega_d \approx 0$\cite{Finn2015}. 
A reduction of the reconnected flux for small rotation was also observed, as an aspect of this resonance phenomenon. 
This screening for small rotation was noted in Refs.~\onlinecite{Liu2012,Fitzpatrick2017}; however, the peaks were not identified as a resonance phenomenon. 
It was also shown in Ref.~\onlinecite{Finn2015} that the net Maxwell torque on the plasma in the tearing layer has a novel form, where the torque is zero near the rotation frequency satisfying the resonance condition, and is maximized for a slightly faster rotation. 
This torque curve leads to the possibility of locking the plasma flow to the mode phase velocity\cite{Finn2015}.
This form of the torque was also observed in the linear resistive MHD simulations (RI regime) by Liu et al.\cite{Liu2012}.

This paper investigates, mostly by means of resistive MHD simulations, the nonlinear behavior in response to a static error field in a rotating plasma with favorable curvature. 
Interestingly, this strong nonlinear effect is observed to emerge in the weakly-driven or `shielded' regime, for which the error fields of interest are not strong enough to lead to magnetic islands that are much larger than the tearing layer width or to significantly decrease the time-averaged plasma rotation rate below $\Omega_0$. 
(In this paper we distinguish screening, the reduction of the linear error field response due to plasma rotation in the presence of real frequency tearing layers, from shielding, a nonlinear effect that prevents locking.) 
In this regime, as the error field $\psiw$ is magnified, the first nonlinear phenomenon to occur appears to be the flattening of the plasma pressure across the island, followed at much larger island width by the usual tearing mode saturation phase. 
The former effect mitigates the GGJ effect, leads to the secondary destabilization of the previously weakly stable TM, and produces magnetic islands that grow larger than the linear tearing layer, indicative of a strong nonlinear behavior even in the shielded state.
%(However, as we discuss later, for more plasma parameters more representative of a tokamak plasma these nonlinear phenomena of pressure and current-flattening may occur in reverse order as $\psiw$ is increased.)
This local pressure flattening effect was first suggested by Finn et al.\cite{Finn2015} and then investigated by Li et al.\cite{Li2017,Li2017_PoP} who, for a Lundquist number of $S=10^9$, showed in a toroidal geometry that the local pressure flattening at the rational surface eliminates the GGJ effect, thereby allowing the full penetration of the RMP. 
However, Li et al.~assumed an \textit{ad-hoc} model for the flattening of the cylindrically symmetric pressure profile because they were performing linear MHD simulations. 
%The linear response for finite $\beta$ to an RMP was also investigated in Ref.\cite{Liu2012} for a circular torus with an aspect ratio of 10. 
An analogous effect related to pressure flattening occurs in two-fluid theory\cite{Biskamp1979,Scott1985}, where unstable drift-tearing modes, with suppressed growth rate and diamagnetic propagation $\omega_{*}$ in their linear phases, stop propagating when the island width $w$ is comparable to a certain critical width $w_{c}$ associated with the flattening of pressure across the island. 
This condition, shown for a Lundquist number of $S=10^5-10^6$ in Ref.~\onlinecite{Scott1985}, is satisfied when the sound-wave frequency $k_{\parallel}c_s \approx k_{\parallel}'(r_t) w c_s$ at the rational surface ($r=r_t$) exceeds the nominal real frequency of the mode, $\omega_r \sim \omega_*$. 
The nonlinear evolution of the mode then proceeds as if the diamagnetic effects were absent\cite{Biskamp1979}. 
%It was observed in Ref.~\onlinecite{Biskamp1979} that the nonlinear evolution of the mode then proceeds as it does without diamagnetic effects.

The simulations in the present paper use a periodic cylinder with an aspect ratio $R/a=10$, and a hollow pressure profile to mimic the toroidal effect of average favorable curvature. 
We focus on modes with $(m,n)=(2,1)$, where modes behave as $e^{im\theta+ikz}$ with $k=-n/R$, yielding a negative Mercier factor $1-q(r_{t})^{2}$, so that in an analogous torus with peaked pressure, favorable curvature would be present. 
Here, $q(r_t)=m/n$ is the safety factor $q(r)=rB_{z}/RB_{\theta}$ at the mode rational surface (MRS) $r=r_{t}$. 
% Also, the minor radius is $a=1$, the major radius $R$ gives an aspect ratio $R/a=10$, . 
The current profile $J_z(r)$ is chosen to be unstable to a spontaneous $(m,n)=(2,1)$ tearing mode in the absence of pressure.
The Lundquist number is set to $S=10^{5}$ for computational expediency and to compare with the results of  Ref.~\onlinecite{Scott1985}; the magnetic Prandtl number is chosen to be $Pr=1$.
The imposed rotation profile $\Omega(r)$ is flat in the interior of the plasma $\Omega(r)= \Omega_0$ with a region of decreasing $\Omega(r)$ near the edge and a no-slip boundary condition $\Omega(a)=0$ at the wall.
In the absence of tearing perturbations, the plasma rotation is maintained by a momentum source whose strength is characterized by $\Omega_0$. 
As will be discussed later, this rotation is assumed to be \textit{poloidal}. 
Lastly, a static error field with $(m,n)=(2,1)$ is applied at the boundary to drive the mode. 

Our first linear MHD simulations have zero plasma rotation and error field: $\Omega_0=\psiw=0$. 
For zero $\beta$ significant growth is observed, as expected. 
As $\beta$ is increased, the growth rate decreases and the first critical value $\beta_m$, corresponding to $\Delta_m$, is crossed above which the modes exhibit the real frequencies associated with the GGJ effect. 
When $\beta$ is raised past the second critical value, $\beta_c$, corresponding to $\Delta_c$, the $(2,1)$ TM is stabilized. 
The linear simulations contain both inertia and viscosity, so that the modes are either in the RI regime or the VR regime\cite{Finn2017} or (most commonly) in the transition range between these regimes, verifying that the Glasser effect is indeed present across a wide range of parameters.
%The stabilization is due to both the positive pressure gradient in the outer region and favorable curvature in the tearing layer\cite{FootNote0}. 
Linear simulations for $\beta$ just above the second critical limit with error field $\psiw$ and rotation $\Omega_0$ show the characteristic form\cite{Finn2015,Finn2017} of the reconnected flux $\psit$ in response to the error field for tearing layers with real frequencies: a double-humped curve with peaks near $\Omega_{0}=\pm\omega_{r}/m$, i.e.~at the phase speed of one of the two modes. 
This response is suppressed around $\Omega_{0}=0$, indicating screening for low frequencies, distinct from the peaked response for $\beta=0$. 
In other words, the reconnected flux is maximized for a plasma that rotates at the phase speed of the mode: $\Omega_0=\pm \omega_r/m$. 
The Maxwell torque across the layer is also computed, and shown to have the form noted in Refs.~\onlinecite{Liu2012,Finn2015,Finn2017}, which features zero-crossings near $\Omega_{0}=\pm\omega_{r}/m$. 
However, as the Maxwell torque is of order $\psit^{2}$, i.e. nonlinear in magnitude, it does not influence the dynamics of these linear simulations. 
This quasilinear effect is calculated only an an additional means to verify the GGJ effect outside the RI regime. 
For increasing values of $\beta$, the mode becomes increasingly damped by the favorable curvature effect and the two peaks of the reconnected flux decrease in amplitude and broaden as a function of $\Omega_0$. 

The nonlinear (NL) resistive MHD simulations shown in this paper investigate the fully nonlinear response to an error field in a plasma with a sufficiently large $\beta$ that the linear tearing mode is weakly stable. Studies are performed with small-to-intermediate error fields and for plasma rotations that are much slower than the Alfv\'en transit frequency. 
For simulations that employ a sufficiently small error field $\psiw$, the fields grow and saturate at a low value described well by linear theory in the time-asymptotic state. 
For simulations employing somewhat larger $\psiw$, but such that the tearing layer is still shielded from the error field, a threshold in the island width $w\approx w_c$ is crossed after the initial transient response to the error field. 
At this time the mode becomes destabilized and grows to a level well beyond its usual response to the error field. 
The flattening of the pressure within the magnetic island eliminates the stabilizing GGJ effect, allowing exponential growth of the mode (and terminating the propagation of the mode in the plasma frame). 
%As in the drift-tearing case, this faster growth and cessation of propagation in the plasma frame are due to flattening of the pressure across the island, which quenches the GGJ effect. 
As a result the mode grows nonlinearly to an amplitude that is determined mainly by the evolution of the current profile \cite{Rutherford1973} and comparable to the saturation of an unstable $\beta=0$ TM. 
In short, the evolution of the magnetic perturbation makes a transition from a driven mode to a spontaneous one when $w>w_c$, and the magnetic islands grow large due to the emergence of the spontaneous mode. 
In this respect the time-asymptotic state appears fully `penetrated' by the error field. 
However, the fact that the time-averaged flow at the MRS exhibits little change suggests that the final state is somewhat shielded from the error field.
%The critical width $w_c$ is empirically determined as the value at which pressure flattening across the island first causes the destabilization and the disappearance of the real frequencies $\pm\omega_r$. 
%The analog of the aforementioned criterion of Ref.~\onlinecite{Scott1985} to determine $w_c$ \textcolor{red}{by substituting $\omega_r$ of GGJ in lieu of $\omega_*$ underestimates the empirically observed $w_c$ by a factor of two-to-three}. This could be because the real frequencies, if they are evident, are seen only during an early transient and typically disappear before such island widths $w \sim w_c$ are encountered.  
%for the suppression of the diamagnetic frequency $\omega_*$ of the drift-tearing mode

The destabilization of the mode due to pressure-flattening is demonstrated by running linear simulations that use as initial condition the $(m,n)=(0,0)$ fields extracted from various times during the NL simulation, initialized with a $(m,n)=(2,1)$ perturbation. 
These `barren' linear runs advance only the $m=2$ mode, with $\psiw=\Omega_0=0$. (The $n=1$ component is calculated \textit{a posteriori} via an FFT)\cite{FootNote5}.
The linear simulations, which use fields extracted from the earlier phase of their parent NL simulation, show linear stability and real (GGJ) frequencies $\pm\omega_r$ , implying $w<w_{c}$. 
The propagation of the real frequencies slows down toward the end of this early phase. 
The simulations that use fields extracted from the following period feature a linear instability and $\omega_r=0$, implying the removal of the GGJ effect and therefore $w_c<w<\delta$. 
The cessation of propagation in the plasma frame occurs almost simultaneously with the destabilization of the mode. 
The growth rate of the mode approximately matches the instantaneous growth rate of the mode from the NL simulations in this phase. 
The pressure inside the magnetic island becomes completely flat by the time of the peak growth rate. 
Beyond this point the growth rate observed from the linear simulations slows down, marking the beginning of the nonlinear saturation phase ($w\approx\delta$) due to the flattening of the current density. 
The simulations that use fields extracted a little beyond this point show linear stability again, but with $\omega_r=0$, implying $w>\delta$ and the full nonlinear saturation of the mode. 
%Note since $\delta$ scales inversely with $S$, it is possible to reverse the ordering $w_{c}<\delta$ for very large $S$, and thereby have the TM saturate before the pressure-flattening event that destabilizes it. However, such a large $S$ is beyond the scope of the present work and necessitates incorporation of additional physical effects such as two-fluid and strong anisotropy. 
%These results, with $w_{c}<\delta$, indicate that the first nonlinear effect encountered as the mode grows is the flattening of the pressure profile and consequent destabilization. 
%The second effect that occurs as the mode grows is flattening of the current profile, leading to the Rutherford phase \cite{Rutherford1973} and the saturation of the mode. 

The interaction of the destabilized spontaneous TM with fields directly driven by the error field, and static in the lab frame, causes an oscillating torque. 
As the spontaneous mode grows, it reaches a stage in which the oscillating torque is large
enough to cause noticeable oscillation in the plasma rotation at $r=r_{t}$, together with oscillations in the phase velocity of the magnetic perturbation at $r=r_{t}$. 
For intermediate values of $\psiw$, the time asymptotic state exhibits these oscillations.
For yet larger values of $\psiw$, the final state is locked (penetrated) in the sense that it has decreased plasma rotation, zero phase velocity for the magnetic perturbations, and a large island. 
%For the parameters we use, the locking effect is the third and last nonlinear phenomenon to be observed {\color{red} and is not the focus of the present work. }
For much larger values of $\psiw$ and the driving rotation $\Omega_{0}$, unambiguous locking is observed, which features a bifurcation with hysteresis\cite{Fitzy1993,Fitzy1998,FinnSovinec1998}. 
We limit the scope of the present work to values of $\psiw$ low enough that locking does not occur; results with locking are deferred to a future publication.

This rest of this manuscript is organized as follows: Section~\ref{sec:CompModel} describes the resistive MHD model as it is implemented in the NIMROD code, and the initial conditions for the simulations.  Section~\ref{sec:LinRuns} discusses the results from the linear simulations, beginning with the demonstration of the emergence of $\pm\omega_r$ and GGJ-stabilization of the spontaneous $(2,1)$ TM, and followed by linear simulations that are initialized from a weakly stable equilibrium with finite plasma rotation, driven by an error field. 
Section~\ref{sec:NLRuns} presents the results of the nonlinear simulations, initialized from the same initial conditions as the linear runs and driven at small-to-intermediate values of $\psiw$. 
%that range from small to intermediate. 
Section~\ref{sec:Discussion} summarizes the findings of this paper and discusses the possible implications of the results.
The Appendix describes a quasilinear model to explain the oscillations observed in the Maxwell torque and other observable quantities. 
%with an unstable spontaneous TM in addition to fields that are driven by the error field, to show that the superposition of the two modes results in the oscillations observed in the Maxwell torque and other observable quantities.

%=====================================================================================================
\section{Computational Model\label{sec:CompModel}}
\begin{figure}
      \centering
      \includegraphics[width=0.5\textwidth]{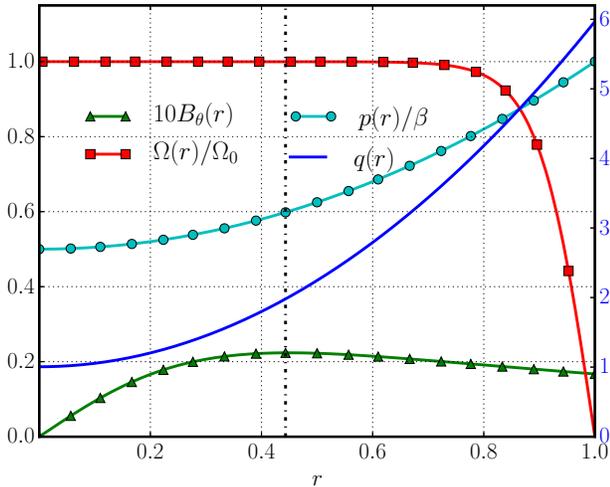} 
     \caption{\label{fig:Eq} The equilibrium profiles of the poloidal magnetic field $B_{\th}$ (green triangles), pressure (cyan circles), poloidal plasma rotation (red squares), and safety factor (solid blue with its scale shown on the right-hand axis) for all of the simulations run with $\beta=0.0016$. The poloidal field $B_{\th}$ is magnified by a factor of 10 to fit on the same scale as the other equilibrium fields.}
\end{figure}

%=====================================================================================================
The simulation geometry is that of a periodic cylinder with an aspect ratio of $R/a=10$,  where $R$ and $a$ are the major and minor radii, respectively.
The equilibrium is prescribed in terms of the safety factor $q(r)$ and pressure profile $p(r)$, from which the magnetic fields are obtained based on the usual magnetohydrodynamic force-balance. 
The pressure profile--shown as the cyan trace with the dots in Fig.~\ref{fig:Eq}--is hollow, with a quadratic dependence on the radial coordinate $r$. 
The hollowness, together with the constant inward curvature in a cylinder, mimics the favorable average curvature of a toroidal plasma with $q>1$. 
This profile has a stabilizing effect both in the inner tearing layer but also in the outer ideal MHD regions, although results discussed in Sec.~\ref{sec:NLRuns} show that the latter effect is negligible. 
The plasma $\beta$ is taken as the ratio of the plasma pressure to magnetic pressure on axis. 
The safety factor profile $q(r)$ is quadratic and spans $1.005\le q(r)\lesssim 6$ (solid blue trace in Fig.~\ref{fig:Eq} with its y-axis appearing on the right). 
The $q=2$ mode rational surface is located at $r\approx 0.43$ for this equilibrium and shown as a vertical dashed-dotted line in Fig.~\ref{fig:Eq}. 
The equilibrium poloidal magnetic field profile is also shown in Fig.~\ref{fig:Eq} (magnified by a factor of ten). 
The toroidal magnetic field profile is omitted from the figure because of the tiny variation ( $\mathcal{O}(a^2/R^2)$) in the toroidal field. 
A nominal rotation rate along the poloidal direction $\Omega(r)=v_{\th}/r$ with a no-slip condition at $r=r_w$ is also imposed (red squares in Fig.~\ref{fig:Eq}). 
This quantity effectively specifies a momentum source that causes a uniform plasma rotation $\Omega=\Omega_0$ across the plasma (except near the wall, where it goes to zero) in the absence of tearing perturbations. 
In a long cylinder, the poloidal component of the mean $E\times B$ velocity $v_{\theta}$ is of order $R/a$ larger than the toroidal component $v_{z}$. In a torus, on the other hand, the flow is predominantly in the toroidal direction due to poloidal flow damping\cite{HassamKulsrud,Morris1996}. 
However, the fact that the angular frequencies $\Omega_{\theta}=v_{\theta}/r$ for the cylindrical case and $\Omega_{\phi}=v_{z}/R$ for the toroidal case are comparable \cite{FootNote1} implies that the Doppler shift for the two cases, $mv_{\theta}^{(cyl)}/r=m\Omega_{\theta}$ and $-nv_{z}^{(tor)}/R=-n\Omega_{\phi}$ are comparable. 
For uniform rotation, only the Doppler shift enters. In this case it suffices to consider poloidal rotation in cylindrical geometry to predict the effects of toroidal rotation in toroidal geometry. 
%; and the remaining physical fields on the right. 
A single-helicity static error field with $(m,n)=(2,1)$ is applied at the boundary, and modes with a helical symmetry of $m/n=2/1$ will be our focus throughout this paper.
The error field is ramped up rapidly, over 10 Alfv\'en toroidal transit times, to a constant value of $m\psiw/a$ where $\psiw$ is the magnitude of the wall flux. 
%---------------------------------------------------------------------------------------
%---------------------------------------------------------------------------------------

The physical system under study is evolved according to the following visco-resistive magnetodhyrodnamics (MHD) equations with an adiabatic closure:
  \begin{align}
        \label{eq:cont}
         \frac{\partial n}{\partial t}+\nabla\cdot(n\mathbf{v}) &=0,\\ %\nabla\cdot D_n\nabla n,\\
         \label{eq:Momentum}
        \rho\left(\frac{\partial\mathbf{v}}{\partial t}+\mathbf{v}\cdot\nabla\mathbf{v}\right) &=\mathbf{J}\times\mathbf{B}-\nabla p-\nabla\cdot\bm{\Pi}, \\
        \label{eq:Ohm}
        \Evec+\vvec\times\Bvec &=\eta \mf{J}, \\
    %    \f{\partial\mathbf{B}}{\partial t} &=-\nabla\times\mathbf{E}+\kappa_{\divB}\nabla\divB,\\
         \label{eq:pres}
          \f{d}{dt}\left(\f{p}{\rho^{\Gamma}}\right)&=0,\mbox{ with } \f{d}{dt}=\f{\part}{\part t}+\vvec\cdot\nabla.
       % n\left(\frac{\partial T}{\partial t}+\mathbf{v}\cdot\nabla T\right) &=-(\Gamma-1) \f{p}{2}\divV,
   \end{align}
Here, $\Bvec$ and $\Evec$ are the magnetic and electric fields. Equation (\ref{eq:Ohm}) is the resistive Ohm's law, which is coupled to the rest of the MHD equations via Faraday's law: $\partial\mathbf{B}/\partial t =-\nabla\times\mathbf{E}$; $\eta$ is the plasma resistivity, which is assumed to be constant and isotropic throughout the plasma.  
The quantities $\Jvec=\nabla\times\mathbf{B}$, $n$, and $\rho=n m_i$ are the current density, the plasma number density, and the mass density, respectively. 
The variable $\vvec$ is the plasma center-of-mass velocity.
%and $T$ represents either the ion and electron plasma temperature, which are assumed to be equal. 
The plasma pressure $p$ is related to the plasma temperature $T$ by the usual equation of state $p=n(T_i + T_e)=2nT$ where equal electron $T_e$ and ion $T_i$ temperatures are assumed. 
The stress tensor employs an isotropic viscosity: $\bm{\Pi}=\nu [\nabla \vvec +(\nabla \vvec)^T-(2/3)\divV]$, where $\nu$ is the kinematic viscosity. 
Lastly, $\Gamma$ is the adiabatic index. 
The Lundquist number is defined as $S=\tau_{R}/\tau_{A}$ where $\tau_{R}=a^{2}/\eta(r=0)$
is the resistive diffusion time and $\tau_A=R/v_A$ is the toroidal Alfv\'en transit time, comparable to the poloidal transit time $a/v_{A\theta}$. 
All the physical quantities are rendered dimensionless by assuming the axial field on axis to be $B_z(r=0)\equiv B_0=1$~T, a minor radius of $a=1$~m, and an initial number density of $n=2\times10^{18}$~m$^{-3}$ for a deuterium plasma, which yield $\Talf\approx 10^{-6}$~secs. 
These choices also result in an error field of magnitude $2\psiw$ for $m=2$. 

Equations (\ref{eq:cont})-(\ref{eq:pres}) are advanced with the NIMROD\cite{NIMRODcode} extended MHD framework. 
NIMROD is a pseudo-spectral code that advances the above resistive MHD system with  semi-implicit\cite{HarnedKerner1985} time stepping. 
Its 2D finite element (FE) structure is used in this case to  discretize the $r-z$ plane of the cylinder, while its finite Fourier series is used to discretize the poloidal circumference of the cylinder. 
Typical resolutions employ 48 radial and 12 axial FE's of third or fourth order, and 3 or 6 poloidal Fourier modes. 
The convergence of the results were checked with finer mesh resolutions as well as with up to 11 Fourier modes. 
The number of Fourier modes is quoted after the dealiasing of the quadratic nonlinearities\cite{FootNote6}. 
Lastly, as NIMROD is not a single helicity code, the required toroidal $n=1$ amplitudes needed for diagnostic purposes were extracted via Fourier transforms. 
 
 %================================================================================================================
 \section{Linear Simulations\label{sec:LinRuns}}
 %================================================================================================================
The simulations in this section advance the linearized version of Eqs.(\ref{eq:cont})--(\ref{eq:pres}) for the $m=2$ mode only. 
The initial equilibrium, described in the previous section, is unstable to the spontaneous $(m,n)=(2,1)$ tearing mode (TM) for $\beta=0$. 
The first series of runs investigate the stabilization of the spontaneous $(2,1)$ TM due to the GGJ effect as a function of the plasma $\beta$, without an error field or plasma rotation. 
As $\beta$ is increased from $\beta=0$ to the value required for marginal stability ( $\beta_c$), the current density profile $J_z(r)$ changes slightly--since we keep the profile $q(r)$ fixed--but not enough to influence linear stability significantly. 
On the other hand, as shown here, the pressure affects the stability properties significantly.

\begin{figure}
   \centering
     \includegraphics[width=0.45\textwidth]{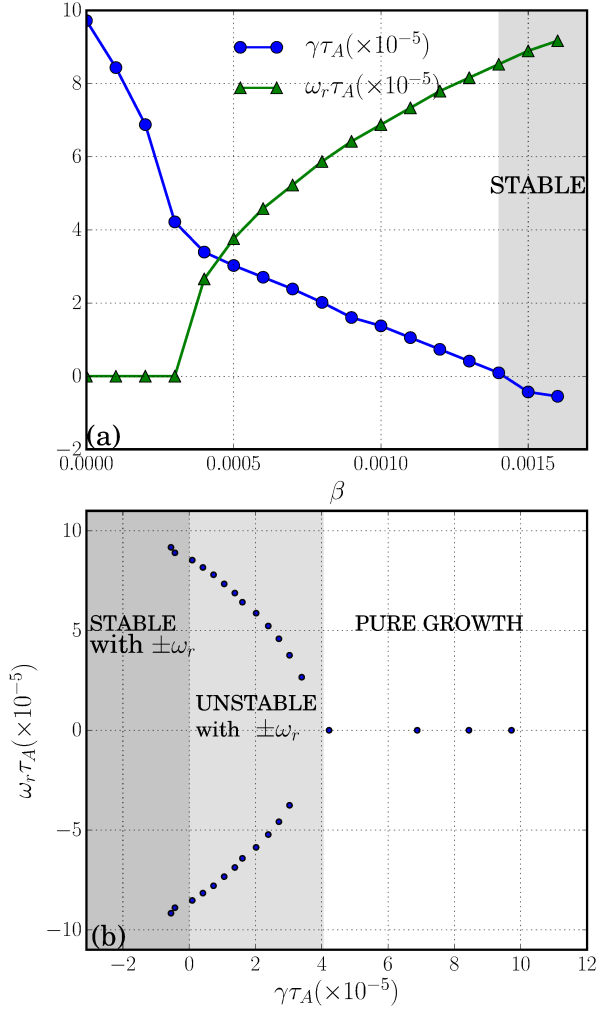}
    \caption{\label{fig:gamma_vs_beta} (a) The normalized growth rate $\gamma\Talf$ and real frequency $\omega_r\Talf$ as functions of the plasma $\beta$ for linear simulations with $S=10^5$. The GGJ oscillations emerge for $\beta>0.0003$ and the $(m,n)=(2,1)$ tearing mode becomes stabilized at $\beta>0.0014$. (b) The locus of roots in the complex plane for the same series of linear simulations.}
\end{figure}

The results from linear NIMROD simulations are shown in Fig.~\ref{fig:gamma_vs_beta}a, which shows the normalized TM growth rate $\gamma\Talf$ (blue dots) and real frequency $\omega_r\Talf$ (green triangles) as functions of $\beta$. 
The propagation of the TM in the plasma frame begins at $\beta =\beta_m \approx 0.0003$, as indicated by the non-zero values of $\omega_r$, and speeds up as $\beta$ increases. 
%which grows as a fractional power of $\beta$, consistent with the linear theory in the RI and the VR regimes  \cite{Glasser1975,Finn2017}.
As $\beta$ is increased, the growth rate (blue trace with circles) steadily decreases with a kink at $\beta_m$ and the mode eventually becomes stabilized ($\gamma<0$) for $\beta>\beta_c=0.0014$. 
%The stabilization is due to both the positive pressure gradient in the outer region and favorable curvature in the tearing layer\cite{FootNote0}. 
% Beyond this point the TM becomes damped with $\gamma<0$. 
Figure \ref{fig:gamma_vs_beta}b portrays the same simulation results as a plot of the locus of roots to highlight the qualitative agreement with the linear theory sketched in Fig.~2b of Ref.~\onlinecite{Finn2015}.
It should be noted that the propagation frequencies are very slow here compared to Alfv\'en scales, reaching $\omega_r\Talf\approx 10^{-4}$ at marginal stability, but are comparable to the typical growth rates found in the RI and VR regimes \cite{Glasser1975,Finn2017}. 

\begin{figure}
   \centering
     \includegraphics[width=0.45\textwidth]{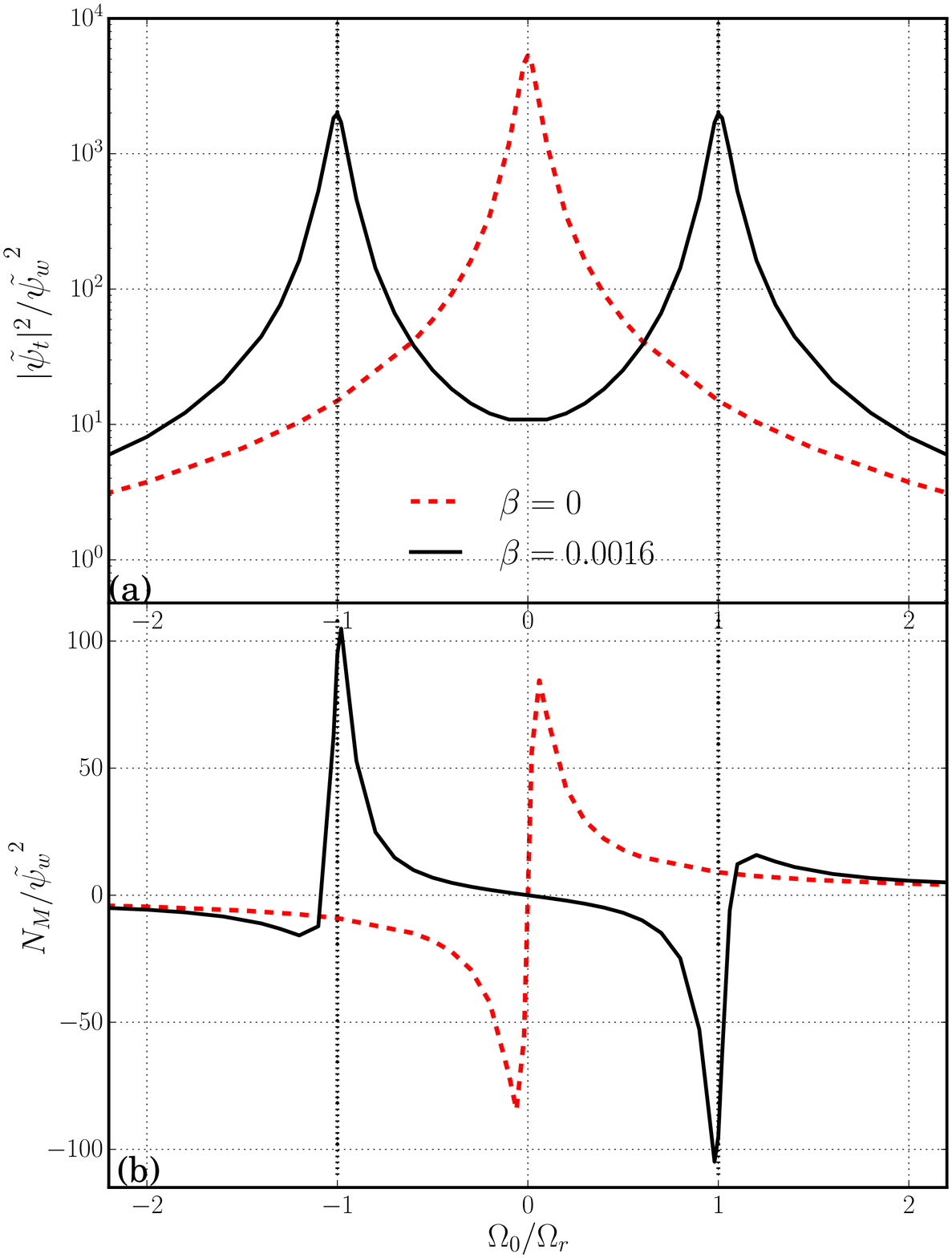}
    \caption{\label{fig:LinPsi_vs_Omega} The (a) magnitude of the reconnected flux (squared) $|\psit|^2$ and (b) the quasilinear Maxwell torque $N_M$ \textit{vs} the normalized poloidal rotation frequency $\Omega_0/\Omega_r$, from linear NIMROD simulations for $\beta=0$ (dashed) and $\beta=0.0016$ (solid). All traces are normalized by $\psiw=5\times10^{-8}$.}
\end{figure}

Linear simulations are also performed to verify the linear response of a weakly damped TM to a $(2,1)$ helical error field for a slowly rotating plasma. 
For these runs and the remainder of all of the simulations featured in this article, we use $\beta=0.0016$ to initialize the simulations in the regime of weak damping. 
This value of $\beta$ corresponds to a poloidal beta of $\beta_p\approx 0.3$ on the MRS. 
For comparison, a series of runs with a force-free equilibrium ($\beta=0$) and a damping rate similar to that of the $\beta=0.0016$ series was also performed. 
The force-free cases, appearing as the dashed red trace in Fig.~\ref{fig:LinPsi_vs_Omega}a. employ a slightly higher $J_z$ on the outboard side of the rational surface ($r>r_t$) than their finite $\beta$ counterparts, to stabilize the spontaneous $(2,1)$ TM in the absence of the GGJ effect.  
Both series of simulations scan a range in the poloidal rotation frequency $\Omega_0$, spanning $-2<\Omega_0/\Omega_r\ < 2$, for a single value of the error field, $2\psiw=10^{-7}$. 
The parameter $\Omega_r\equiv v_{ph}/r$ is the poloidal phase velocity of the mode, related to the real frequency of the mode by  $\omega_r=k v_{ph}$ with $k= m/r$ or $\omega_r=m\Omega_r$. 
Thus, the phase velocity for $m=2$ is half of the real frequency of the mode.

The linear results with error fields are run to a time-asymptotic state that is attained when the magnitude of the reconnected flux at the MRS $|\psit|$ becomes steady.
This quantity is calculated from the $(2,1)$ component of the radial magnetic field $b_r$ at MRS via: $\psit=i r_t b_r(r=r_t)/m$. 
Figure \ref{fig:LinPsi_vs_Omega}a shows $|\tilde{\psi}_t|^2$, normalized by $\psiw^2$, as a function of $\Omega_0$ for $\beta=0$ (dashed red trace) and $\beta=0.0016$ (solid black trace) at $2\psiw=10^{-7}$. 
The values on the horizontal axis are normalized to the GGJ rotation frequency $\Omega_r$ at $\beta=0.0016$. 
The reconnected flux at the MRS $|\psit|$ for $\beta=0$ exhibits the usual Lorentzian shape centered at $\Omega_0=0$, an indication of the VR tearing regime\cite{FootNote2}. 
The reconnected flux traces for $\beta=0.0016$ feature strong screening for slow $\Omega_0$ and two symmetric peaks located approximately at $\Omega_0=\pm \Omega_r$, consistent with the quasilinear theory\cite{Finn2015}. 
Each peak corresponds to the Doppler-shifting of either the forward or backward propagating mode to zero frequency in the lab frame, which results in the observed resonance with the static error field. 
Away from this resonance, the transients in the linear simulations exhibit two frequencies, $\Omega_0\pm\Omega_r$, one representing a fast beat and the other a slow one, depending on the relative signs of $\Omega_0$ and $\Omega_r$. 
Right on resonance, the slow frequency is zero and the fast frequency is  $2\Omega_r$. As discussed in Refs.~\onlinecite{Finn2015,Finn2017}, the actual maxima of $|\psit|$ lie slightly to right/left of $\Omega_r$ for positive/negative $\Omega_0$. 
This offset is negligible for the weakly damped case but grows as $\beta$ increases.
% The reconnected flux for the case closest to marginal stability  ($\beta=0.0112$) is sharp with a similar peak amplitude and damping rate to those from $\beta=0$.
As $\beta$ is raised even further and the mode becomes more strongly damped, $|\psit|$ decreases in magnitude and the response curve depicted in Fig.~\ref{fig:LinPsi_vs_Omega}a broadens. 
%As expected from linear and quasilinear theories, the normalized flux and Maxwell torque from the scan with $\psiw=5\times10^{-8}$ are identical to those for $2\psiw=10^{-7}$. I still think that, in the interest of space, we could leave this out.

The Maxwell torque $N_M$ exerted on the tearing layer in the poloidal direction is proportional to $\int r dr \langle b_r j_z^* - b_z j_r^* \rangle_{\th}+ \mbox{c.c.}$, where $b_r$ etc. represent the $(m,n)=(2,1)$ Fourier amplitudes. 
These quasilinearly computed torques are in the context of linear theory in this section, i.e.~they do not affect the plasma rotation. 
The torques for zero and finite $\beta$ are plotted as functions of $\Omega_0/\Omega_r$ in Fig.~\ref{fig:LinPsi_vs_Omega}b. 
The torque curve for finite $\beta$ differs from that of $\beta=0$ in three major ways: (1) the zero-crossings for finite $\beta$ shift from $\Omega_0=0$ to the vicinity of $\pm\Omega_r$, (2) the extrema in the torque curve also occur near $\Omega_0=\pm\Omega_r$, in the immediate vicinity of (but beyond) the zero-crossings, and (3) the torque reverses over $\Omega_0 \lesssim |\Omega_r|$. 
For a mode exactly at the stability boundary ($\gamma=0$), the zero-crossings in $N_M$ should correspond exactly to the maxima of $|\tilde{\psi}_t|$, which should lie at $\pm \Omega_r$ \cite{Finn2015}. 
Similar to the reconnected flux plotted in Fig.~\ref{fig:LinPsi_vs_Omega}a, the torque curves are sharp near marginal stability, and decrease in amplitude and broaden for values of $\beta$ that are much larger than that at marginal stability ($\beta>0.0014$). 

The forms of the reconnected flux and Maxwell torque as functions of plasma rotation, emerging from the present linear simulations, are consistent with the quasilinear theory of tearing and error-field penetration in the presence of real frequencies that are driven by pressure gradient and magnetic curvature\cite{Finn2015,Finn2017}. 
The above results show that these effects are present in a tearing regime other than the RI regime since these results involve ion inertia and viscosity, i.e.~are between the RI and VR regimes. Indeed, $|\psit|$ as a function of $\Omega_0$ for $\beta=0$, in Fig.~\ref{fig:LinPsi_vs_Omega}a, has qualitative features of a VR response, and except for pressure and the small change in the profile $J_z(r)$ utilized to stabilize the spontaneous TM for $\beta=0$, the parameters of the two cases are the same.

%==============================================================================
\section{Nonlinear Simulations\label{sec:NLRuns}}
\begin{figure}
    \centering
    \includegraphics[width=0.5\textwidth]{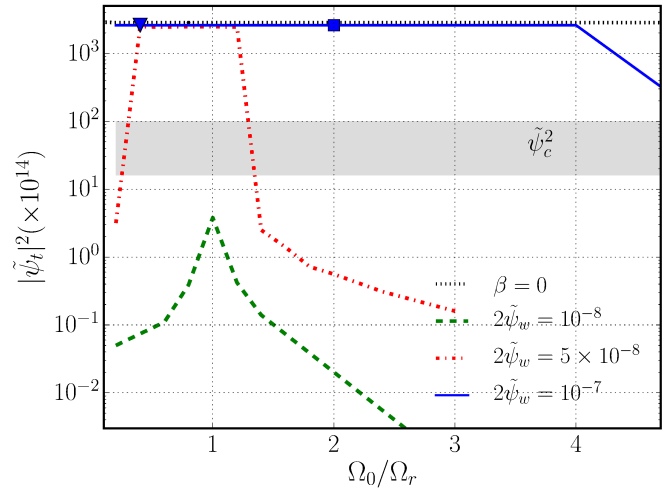} 
     \caption{\label{fig:Psit_vs_Omega0_NL} Norm squared of the reconnected flux at the MRS $\psit$ vs the (normalized) initial rotation frequency $\Omega_0$ for different values of the error field $\psiw$ at $\beta=0.0016$ and $S=10^5$. The scale is set to a nominal wall flux of magnitude $\psiw=10^{-7}$.
     The gray band indicates the range for the empirical values of the critical mode amplitude $\psic$ that signals the destabilization of the spontaneous $(2,1)$ TM. For comparison, a series of NL $\beta=0$ runs (dotted horizontal line), unstable to the spontaneous $(2,1)$ TM and driven at $2\psiw=10^{-7}$, is also plotted.}
\end{figure}

%==============================================================================
A series of nonlinear (NL) simulations scanning over different initial rotation frequencies $\Omega_0$ was conducted for different magnitudes of the error field $\psiw$, once again for an equilibrium that is weakly stable to the $(2,1)$ TM for $\beta=0.0016$. 
The two parameters $\Omega_0$ and $\psiw$ are the main \textit{control parameters} for the NL effects we study.
The range of rotation frequencies were chosen to be fairly low: $\Omega_0\Talf=10^{-5}-10^{-4}$, in the vicinity of $\Omega_r$. 
Error fields of relatively low magnitude, $10^{-8}\le2\psiw\le5\times10^{-7}$, were employed to remain in the shielded or unlocked regime. 
The main nonlinear effects are characterized by measuring two \textit{order parameters}: the time-asymptotic magnitude of the reconnected flux at the mode rational surface (MRS), $|\psit|$, and the plasma rotation $\Omega_t$ there. 
The latter quantity is a poloidal and toroidal average of the poloidal rotation at the MRS. 
The Lundquist number $S$ is $10^5$. 
Studies employing larger values of the two control parameters, as well as larger values for $S$ are deferred to a future publication. 
The magnitude of the reconnected flux $|\psit|$ as a function of $\Omega_0$ for several values of $\psiw$ is plotted in Fig.~\ref{fig:Psit_vs_Omega0_NL}. 
%These values of $|\psit|$ are normalized to \textcolor{blue}{a nominal value of the wall flux, $2\psiw=10^{-7}$ in the figure to emphasize the fact that the tearing mode saturates at the same amplitude in the regime of interest, regardless of $\psiw$. }
The response at $2\psiw=10^{-8}$ appears to be linear: it is comparable to $|\psiw|$ with a sharp peak centered around the mode's phase velocity, $\Omega_0/\Omega_r=1$, very similar to the linear results shown in Fig.~\ref{fig:LinPsi_vs_Omega}. 
Dialing up the error field slightly ($2\psiw\ge 2\times10^{-8}$) evinces a dramatic departure from the linear behavior: at $2\psiw=5\times10^{-8}$ the resonant response of the reconnected flux to the error field broadens, becoming flat over a small but noticeable range in $\Omega_0$. 
For $2\psiw=10^{-7}$, the $|\psit|$ curve broadens further with the region of the flat resonant response spanning a range $-4\Omega_r\lesssim \Omega_0\lesssim 4\Omega_r$. 
In other words, the region of strong nonlinear response widens--while $|\psit|$ remains fixed--as $\psiw$ is raised, reaching plasma rotation frequencies that are much faster than $\Omega_r$. 
Over this flat response regime, $|\psit|$ dwarfs $\psiw$ with $|\psit|/\psiw= 200$ for $2\psiw=5\times10^{-8}$ and $|\psit|/\psiw= 100$ for $2\psiw=10^{-7}$, respectively. 
This is an amplification factor that far exceeds the levels from the linear regime and those reported by Refs.~\onlinecite{Li2017,Li2017_PoP}. 
The reason why this amplification factor is greater for the lower $\psiw$ is that the time-asymptotic amplitude of the mode is the same as indicated by Fig.~\ref{fig:Psit_vs_Omega0_NL}, regardless of $\psiw$ in the regime of interest. 
Increasing the error field beyond $2\psiw\approx3\times10^{-7}$ leads to the locking of the plasma flow to zero frequency because the Maxwell torque, which scales as $\psit^2$, continues to grow. 
In this paper, we limit our investigation to `unlocked' or shielded cases, i.e. those driven with $2\psiw\lesssim10^{-7}$, as the phenomenon of locking is not our focus.  We note here that the results of this paper suggest that phenomena like locking, having large islands, behave as if $\beta=0$. That is, the locking process appears not to be influenced by pressure.
%In fact, locking already begins to creep in at $2\psiw\approx10^{-7}$ for $\Omega_0\le10$~rads/sec. 

In order to understand the mode amplification beyond the error-field drive, the $\beta=0.0016$ scan driven with $2\psiw=10^{-7}$ was repeated for $\beta=0$, i.e. with an initial equilibrium that uses the same $q(r)$ as the finite $\beta$ equilibrium, but is now unstable to the spontaneous $(2,1)$ TM because the stabilizing GGJ effect is removed. 
The reconnected flux $|\psit|$ from the $\beta=0$ case is plotted as horizontal dots in Fig.~\ref{fig:Psit_vs_Omega0_NL}, showing that the mode amplitude for finite $\beta$ attains nearly the same level as that of a $\beta=0$ unstable TM. (The response from the $\beta=0$ scan is independent of $\Omega_0$ because the mode is intrinsically unstable and therefore, indifferent to the shielding by the error field for fast flows, which simply convect the growing mode.) 
This comparison further corroborates the existence of a mechanism that causes growth of a weakly-stable driven finite-$\beta$ TM as though it is an unstable spontaneous $\beta=0$ TM. 

Figure~\ref{fig:Psit_vs_Omega0_NL} applies to cases that are initially weakly stable because of finite $\beta$. 
For very small values of $\psiw$, these cases show linear saturation similar to that discussed in Sec.~\ref{sec:LinRuns}. 
For larger values of $\psiw$ an amplification of ${\psit}$ well beyond the linear response occurs. 
For these cases, one can empirically define a critical amplitude of the reconnected flux, $\psic$, which signals this amplification. 
This critical amplitude is shown as the gray shaded area in Fig.~\ref{fig:Psit_vs_Omega0_NL} in dimensionless form. 
The width of the area represents the variation in the values of $\psic$ extracted from the simulations. 
% Observe while $|\psit|/\psiw$ seemingly exceeds $\psic/\psiw$ for $2\psiw\approx10^{-8}$, the peak amplitude from this series is approximately $|\psit|=2\times10^{-7}$ always stays below $\psic$. 
One can re-express $\psic$ in terms of a critical island width $w_c$ by using the relation between the (full) island  width $w$ and $|\psit|$: $w=4\sqrt{|\tilde{\chi}_t|/\chi''_0(r_t)}$ where the perturbed helical flux at the MRS is $\tilde{\chi}_t\approx m\psit$ and the curvature of the equilibrium helical flux yields $\chi''_0(r_t)\approx -r_t B_0 k'_{\parallel}(r_t)$ in the reduced MHD limit. 
Thus, the critical island width $w_c$ associated with $\psic$ is:
\be
\label{eq:Wcr}
   w_c = 4\sqrt{\left | \f{m\psic}{r_t B_0 k'_{\parallel}(r_t)}\right |}.
\ee

For most cases under study here, $\psic\approx 5\times10^{-7}$, which corresponds to a critical island width $w_c=0.013$ according to Eq.~(\ref{eq:Wcr}). 
This empirical value for the critical width is approximately three times the analog of the result of Ref.\onlinecite{Scott1985} that is based on $\omega_*$ (further discussed in Sec.~\ref{sec:Discussion}). 
The critical mode amplitude $\psic$ approximately corresponds to an error field of magnitude $2\psiw=2\times 10^{-8}$. 
This value is consistent with the results of additional NL simulations (not shown in Fig.~\ref{fig:Psit_vs_Omega0_NL}) that indicate that the transition from the linear to NL behavior described above indeed occurs for $2\psiw\ge2\times 10^{-8}$. 
Critical amplitudes as large as $10^{-6}$ have been observed in cases where $\Omega_0\approx\Omega_r$. 
It is shown in the following paragraphs how $\psic$ is determined.  
%from the NL simulations that exhibit the destabilization of the spontaneous $(2,1)$ TM due to the flattening of the equilibrium pressure across a growing magnetic island. 

\begin{figure}
    \centering
     \includegraphics[width=0.5\textwidth]{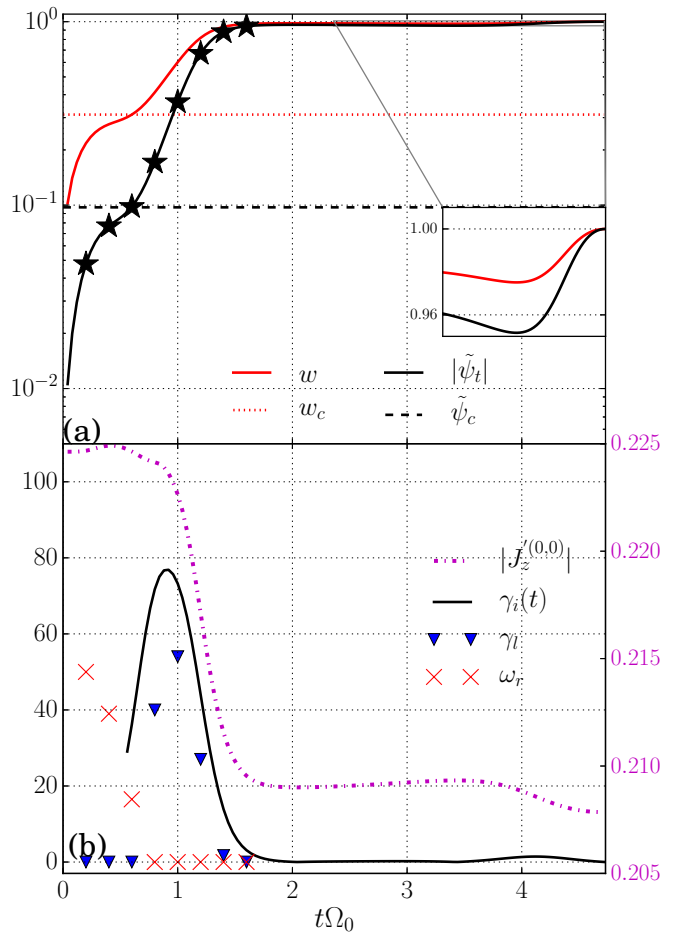}
     \caption{\label{fig:Psit_width_vs_Omega20_NL} (a) Temporal traces of the magnitude of the reconnected flux at the MRS $|\psit|$ (solid trace with stars), island width $w$ (solid red trace) for a nonlinear simulation driven with $2\psiw=10^{-7}$ and $\Omega_0=2\Omega_r/5$, marked by the blue triangle in Fig.~\ref{fig:Psit_vs_Omega0_NL}. (b) Each real frequency ($\times$ marks) and linear growth rate (triangles) corresponds to data from a single linear restart that uses as its initial equilibrium the $(0,0)$ fields extracted from the times marked by stars in (a). The gradient of the equilibrium current density $|J_z^{'(0,0)}(r_t)|$ (dashed magenta) is also tracked in the parent nonlinear simulation, with the vertical axis on the right providing its scale. }
\end{figure}

Figure~\ref{fig:Psit_width_vs_Omega20_NL}a shows the temporal traces of the reconnected flux at the MRS $|\psit|$ and island width $w$ on a log scale, for a case driven with $\Omega_0=20$ and $2\psiw=10^{-7}$ that is marked by the blue triangle in Fig.~\ref{fig:Psit_vs_Omega0_NL}. 
%Since $w\propto \sqrt{|\psit|}$, the evolution of $w$ mimics that of $|\psit|$, except it is suppressed by a factor of the square root. 
Both traces are normalized by their respective maximum value, and the simulation time is normalized by the initial rotation frequency: $t\rightarrow t\Omega_0$. 
The figure shows an initial growth stage due to the error field: saturate begins at $t\approx0.03$ in a manner similar to saturation of a weakly stable mode driven by a weak error field. 
This saturation phase terminates at $t\approx0.06$ and is soon followed by second and longer growth period where $|\psit|$ is amplified ten-fold due to an apparent linear instability. During this phase the width $w$ roughly triples in size. 
We identify the $|\psit|$ at the beginning of this linear growth period as $\psic$ and mark it by the dashed horizontal line, which corresponds to $\psic=5\times10^{-7}$ and $w_c=0.013$, where the latter quantity is represented by the dotted red horizontal line.  
%This value for $\psic$ is several factors smaller than the maximum amplitude observed from the linear simulations driven by the same $\psiw$. 
Our hypothesis is that the observed growth phase is caused by the quenching of the GGJ effect due to the flattening of pressure within the magnetic island, which results in the cessation of the propagation of the mode $\omega_r\rightarrow0$ in the plasma frame and destabilization of the spontaneous $(2,1)$ TM. 
Following this growth phase, the mode enters a second and final saturation phase where the island size becomes comparable to the linear layer width $\delta\approx0.03$ (VR theory yields $\delta=0.035$ for the present parameters). 
%marked in time by an abrupt drop in $|J_z^{'(0,0)}(r_t)|$.
This is the usual Rutherford phase associated with the flattening of the current profile, signaling the beginning of nonlinear saturation. 
In the ensuing time-asymptotic stage, $|\psit|$ and $w$ are seen to undergo oscillations of small-amplitude, as shown in the inset of Fig.~\ref{fig:Psit_width_vs_Omega20_NL}a. 
We elaborate on the nature of these oscillations further below. 

The destabilization of the intrinsic $(2,1)$ TM due to the pressure-flattening is demonstrated by performing linear simulations that use as initial conditions the $(m,n)=(0,0)$ profiles extracted from various times in a particular parent NL simulation, and using an arbitrary initial $(m,n)=(2,1)$ perturbation to seed the mode. 
These simulations evolve only the $m=2$ mode\cite{FootNote5} with $\psi_w=0$ and $\Omega_0=0$. 
As such, they can be thought of the `barren' linear restarts of their parent NL simulation. 
The chosen restart times from the NL runs correspond to (I) several instants when $w<w_c$ (or $|\psit|<\psic)$, (II) others when $w_c<w<\delta$, and (III) finally a few points from the nonlinear saturation phase, signaled by when $w>\delta$. 
These times are marked by stars on the $|\psit|$ trace for the particular example shown in Fig.~\ref{fig:Psit_width_vs_Omega20_NL}a and the results of the corresponding linear restarts are shown in Fig.~\ref{fig:Psit_width_vs_Omega20_NL}b. 
The latter figure plots the observed linear growth rates $\gamma_l$ (blue triangles) and real frequencies $\omega_r$ (red crosses) as well as the logarithmic derivative of $|\psit|$ with respect to time from the parent NL run, to provide an instantaneous growth rate $\gamma_i(t)$. 
The absolute value of the gradient of the $(0,0)$ component of $J_z$ at the MRS, $|J_z^{'(0,0)}(r=r_t)|$ (magenta dashed and dotted trace), is also tracked as a proxy for the nonlinear saturation due to the flattening of the current. 
The linear growth rate $\gamma_l$ is extracted from the temporal evolution of the $m=2$ kinetic energy and verified by the temporal evolution of $|\psit|$ from the same linear simulation. 
The instantaneous growth rate $\gamma_i(t)$ is plotted only for times when $w>w_c$. 
We use the outcome of these linear simulations, specifically whether they yield $\gamma_l=0$ or $\gamma_l>0$ to further constrain the value of $\psic$ that is initially estimated from the temporal evolution of $|\psit|$ shown in Fig.~\ref{fig:Psit_width_vs_Omega20_NL}a. 
%------------------------------------------------------------------------
\begin{figure}
    \centering
    \includegraphics[width=0.5\textwidth]{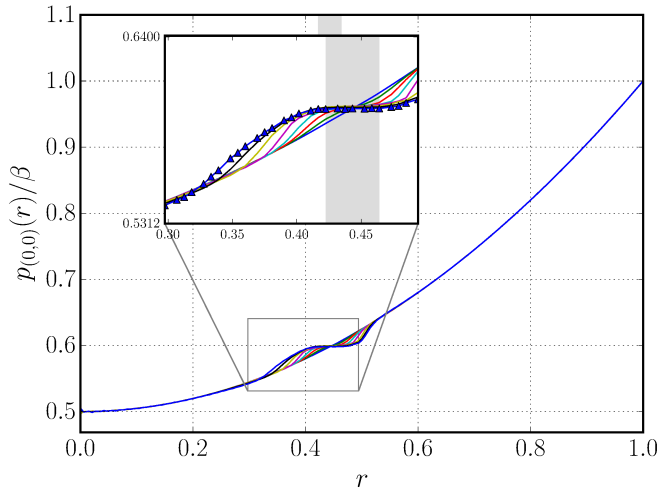}
     \caption{\label{fig:P00_Profs_Om20Br1eM7} Profiles of the $(0,0)$ pressure normalized by $\beta$ corresponding to the times marked by the stars in Fig.~\ref{fig:Psit_width_vs_Omega20_NL}a. The shaded region represents the radial extent of the magnetic island at saturation. The blue trace with the triangles represents the final pressure during the secondary instability stage.  The flattening of the initial pressure profile across the island and beyond is evident as time progresses. }
\end{figure}
%------------------------------------------------------------------------

\begin{figure}
  \centering
  \includegraphics[width=.5\textwidth]{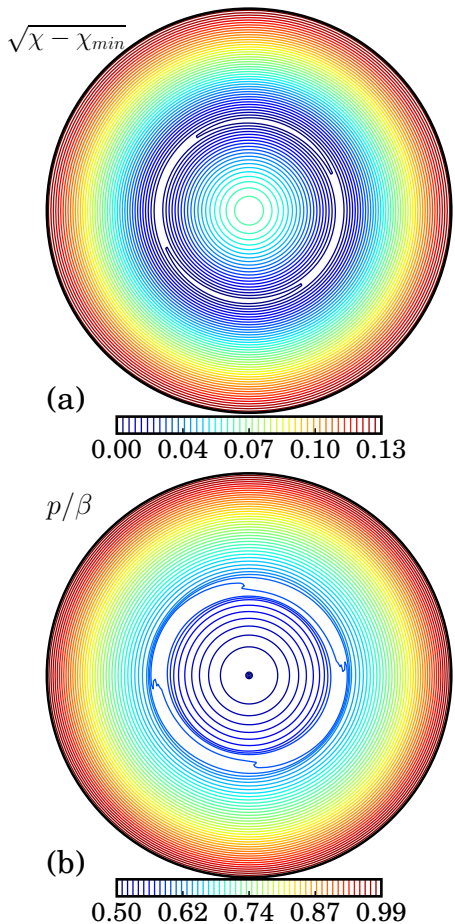} 
   \caption{\label{fig:SqrtChiandPres_cont_Om20Br1eM7}Contours of (a) $ \sqrt{\delta \chi}$, showing the $2/1$ magnetic island and (b) $(0,0)$ pressure at $t\Omega_0\approx 1$, in the middle of the destabilization period when the linear growth rate is at its maximum. After this time, the island continues to grow, albeit at a slower rate, until the saturation of the mode and doubles its depicted size here. }
\end{figure}

For the initial stage where the error field drives the mode, the intrinsic TM is stable, as indicated by data points with $\gamma_l=0$, while the real frequency $\omega_r$ starts gradually slowing down, consistent with the beginnings of the flattening of the $(0,0)$ pressure profile, represented by the solid blue and green traces in the inset of Fig.~\ref{fig:P00_Profs_Om20Br1eM7}. 
The next stage, marked by $w>w_c$, while $w<\delta$, exhibits data points with $\gamma_l>0$, signaling the expected destabilization of the $(2,1)$ TM associated with the pressure-flattening. 
During this period, the pressure is observed to flatten across the magnetic island as can be seen by the red and cyan traces in the inset of Fig.~\ref{fig:P00_Profs_Om20Br1eM7}. 
As the linear growth rate $\gamma_l$ rapidly rises, $\omega_r$ just as rapidly drops to zero, unfolding the full aftermath of the termination of the GGJ effect. 
The time at which the mode becomes destabilized ($\gamma_i>0$) leads that at which real frequencies vanish $\omega_r=0$, but only marginally. 
It is seen from the figure that $\gamma_l$ tracks $\gamma_i(t)$ in time closely, but underestimates it somewhat. (Since $\gamma_l$ is based on only the $(m,n)=(0,0)$ Fourier component, it is not expected to equal $\gamma_i(t)$ exactly.)
The peaks of $\gamma_l$ and of $\gamma_i(t)$ occur at $t\approx0.16$ for the particular case shown in Fig.~\ref{fig:Psit_width_vs_Omega20_NL}. 
By this time as indicated by the blue trace with the triangles in the inset of Fig.~\ref{fig:P00_Profs_Om20Br1eM7}, the pressure has completely flattened across and beyond the magnetic island, demarcated by the shaded region in Fig.~\ref{fig:P00_Profs_Om20Br1eM7}. 
%In fact, the region of flat pressure spans an area that is nearly twice the island width $w$, demarcated by the gray shaded region in Fig.~\ref{fig:P00_Profs_Om20Br1eM7}. 
This is also evident from Fig.~\ref{fig:SqrtChiandPres_cont_Om20Br1eM7}, which provides a side-by-side comparison of the contours of the (a) $\sqrt{\delta \chi}=\sqrt{\chi(r,\theta)-\chi(r=0)}$, where $\chi$ is the helical flux $mA_z-krA_{\theta}\approx mA_z$, indicating the magnetic island and (b) $(0,0)$ pressure at the time of peak $\gamma$. 
The reason why the region of flattened pressure is broader than the magnetic island width appears to be related to the fact that the flow structure associated with the reconnection in a saturated mode extends outside the island\cite{FinnSovinec1998}. 
The mode still grows after the maximum growth rate is attained, but at a decreasing rate, in conjunction with the flattening of current across the magnetic island, indicated by a sudden drop of 15\% in the gradient of $J_z^{(0,0)}$. 
This marks in time the Rutherford phase ($w\approx\delta$), 
the beginning of the nonlinear saturation of the mode due to flattening of the equilibrium current. 
The linear growth ceases at $t\gtrsim0.23$ and the mode attains its steady-state amplitude shortly after that, with a final time-asymptotic island width of $w=0.04$. 
%------------------------------------------------------------------------
\begin{figure}
    \centering
    \includegraphics[width=0.5\textwidth]{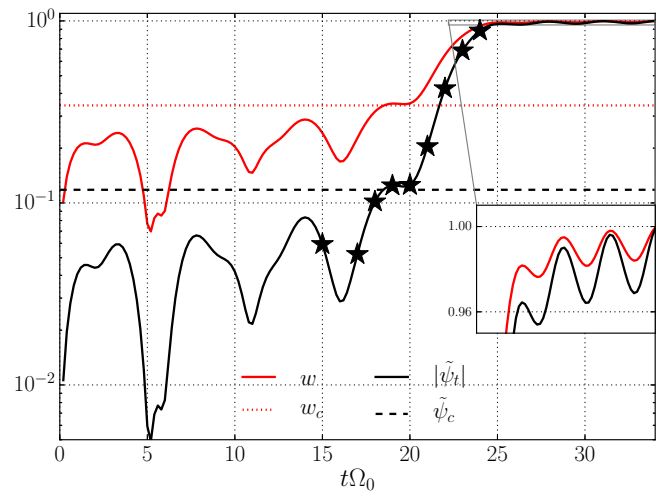} 
     \caption{\label{fig:Psit_width_vs_Omega100_NL} Temporal traces of the magnitude of the reconnected flux at the MRS $|\psit|$ (solid trace with stars), island width $w$ (solid red trace) for a nonlinear simulation driven with  $2\psiw=10^{-7}$ and $\Omega_0=2\Omega_r$, marked by the blue square in Fig.~\ref{fig:Psit_vs_Omega0_NL}. The critical values $w_c$ and $\psic$ are labeled as in Fig.~\ref{fig:Psit_width_vs_Omega20_NL}.
     The transient following the initial linear response to the error field and prior to the destabilization of the spontaneous mode lasts much longer for this case.}
\end{figure}
%------------------------------------------------------------------------

The characteristics described in the preceding paragraphs apply to all cases exhibiting the secondary destabilization of the intrinsic $(2,1)$ TM due to the flattening of pressure across the magnetic island: The duration of the linear instability associated with the pressure-flattening is approximately the same for all cases, as is the value of $\psic$. 
There is however a notable difference related to the length of the transient prior to the destabilization of the mode: as $\Omega_0$ is raised further beyond $\Omega_r$, a long-lasting transient appears, following the initial response to the error field as shown in Fig.~\ref{fig:Psit_width_vs_Omega100_NL}a. 
In fact, faster $\Omega_0$ is associated with longer the duration of the transient: e.g. the transient for $\Omega_0\lesssim\Omega_r$ lasts less than one cycle $2 \pi /\Omega_0$ while it can last as long as 80 cycles for $\Omega_0\gtrsim 4\Omega_r$. 
If the poloidal rotation frequency $\Omega_0$ is not too large, this transient eventually succumbs to the destabilization of the mode identified above, after which the NL simulation progresses in a nearly identical fashion to the other cases driven with a slower $\Omega_0$, as shown in Fig.~\ref{fig:Psit_width_vs_Omega100_NL}b. On the other hand, if $\Omega_0$ is so fast as to shield the tearing layer from the error field, the transient stage becomes permanent. 
Again, note the small amplitude-oscillations that appear in the temporal traces of $|\psit|$ and $w$ in Fig.~\ref{fig:Psit_width_vs_Omega100_NL}a during the time-asymptotic stage, which we visit next.
%--------------------------------------------------------------------------------------------------------
\begin{figure}
    \centering
    \includegraphics[width=.4\textwidth]{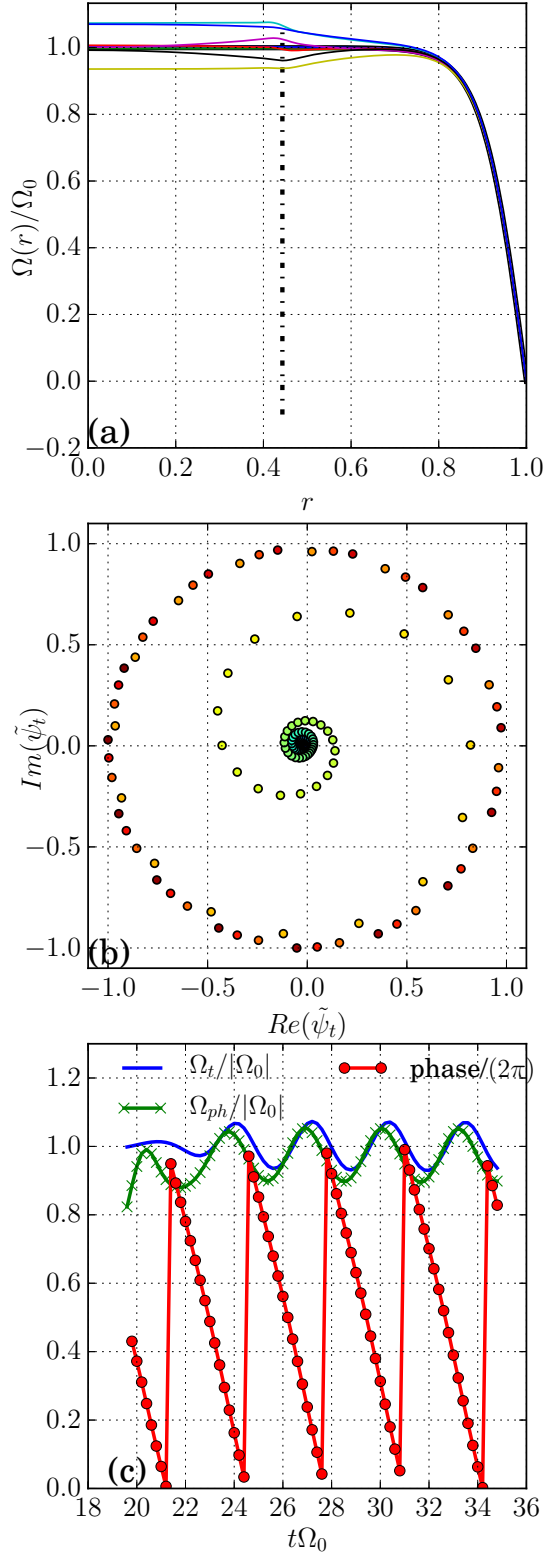} 
    \caption{\label{fig:Oscillations} The panels are from the NL simulation of Fig.~\ref{fig:Psit_width_vs_Omega100_NL}:  (a) Traces of the normalized poloidal rotation profile showing the fluctuations over an area extending from the axis to nearly the wall. (b) A phase plot of the tearing mode: the spectrum of colors from blue to red represent the chronological order of the phase in the simulation; blue corresponds to earlier times and red to later times. (c) Temporal traces of the phase and phase velocity of the mode as well as the fluid rotation frequency at the mode rational surface, $\Omega_t$. Only the post-saturation phase of the simulation is shown.}
\end{figure}
%--------------------------------------------------------------------------------------------------------

An observable consequence of the pressure-flattening is the emergence of persistent oscillations at a frequency of $\bar{\Omega}_t$ in $w$, $|\psit|$, the fluid rotation frequency at the MRS $\Omega_t$, and the phase velocity of the mode $\Omega_{ph}$; where the factor of 2 comes from $m=2$. 
The quantity $\bar{\Omega}_t$ is the time-averaged fluid rotation frequency at the MRS and the nonlinear phase velocity $\Omega_{ph}$ should not be confused with the phase velocity from linear theory, $\Omega_r$, that was introduced in Section~\ref{sec:LinRuns}. 
The oscillations in question are quite distinct from the GGJ oscillations observed in the transient that precede the secondary destabilization. 
They appear long after the GGJ oscillations terminate, as part of the nonlinear behavior of the mode for all cases that feature the destabilization of the $(2,1)$ TM, and are essentially identical to the oscillations observed for $\beta=0$ cases with an unstable spontaneous mode. 
An illustration of the said oscillations is provided in Fig.~\ref{fig:Oscillations} which corresponds to a NL simuilation driven with $2\psiw=10^{-7}$ and $\Omega_0=2\Omega_r$, marked by the blue square in Fig.~\ref{fig:Psit_vs_Omega0_NL}
They have a global effect on the plasma flow, causing fluctuations in the fluid rotation frequency throughout the volume (Fig.~\ref{fig:Oscillations}a). 
These fluctuations are most prominent for very slowly-rotating cases where $\bar{\Omega}_t$ is seen to slow down, as much as 25\% with respect to $\Omega_0$ for the case with $\Omega_0=2\Omega_r/5$ (Fig.~\ref{fig:Psit_width_vs_Omega20_NL}). 
The phase velocity of the mode $\Omega_{ph}$ also oscillates about $\Omega_0$ (Fig.~\ref{fig:Oscillations}c).
In fact, $\Omega_{ph}$ and $\Omega_t$ oscillate nearly in phase, with $\Omega_{ph}$ slightly leading $\Omega_t$ ( Fig.~\ref{fig:Oscillations}c). 
As the insets of Figs.~\ref{fig:Psit_width_vs_Omega20_NL} and \ref{fig:Psit_width_vs_Omega100_NL} indicate, magnetic islands also undergo cycles of expansion and contraction, associated with the oscillations, although the amplitude of these oscillations is quite small. 
A phase plot of the tearing mode at the MRS is also included, in Fig.~\ref{fig:Oscillations}b. 
Here, blue dots correspond to earlier times in the simulations and red dots to later times. 
Note the non-uniform spacing between the dots from the later stage, indicative of the non-uniform phase speed of the mode. 
Also note the very small changes in the radius (representing $|\psit|$) during the asymptotic-phase, which is consistent with the oscillations of very small amplitude observed in the time-asymptotic island width. 

\begin{figure}
    \centering
    \includegraphics[width=0.5\textwidth]{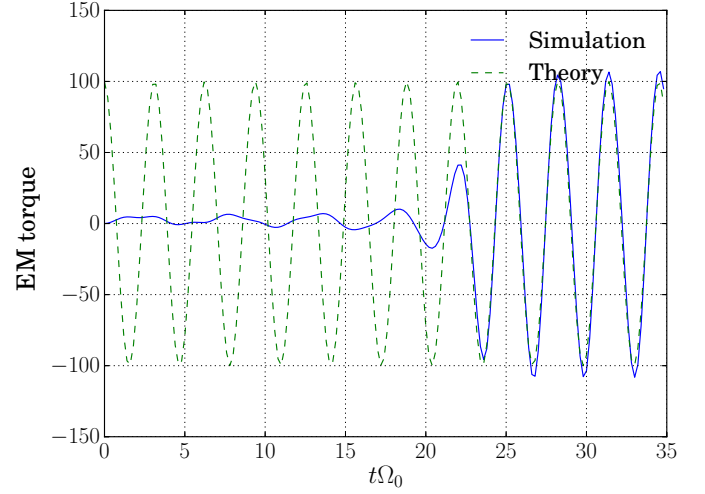} 
     \caption{\label{fig:Torques_vs_time} Temporal traces of the quasilinear Maxwell torque from the nonlinear simulation featured in Figs.~\ref{fig:Psit_width_vs_Omega100_NL} and \ref{fig:Oscillations} (solid trace), and from Eq.~(\ref{eq:Maxwell-torque}) of the analytic theory presented in the Appendix (dashed trace), with $\hat{\psi}_t$ given by the saturated value at $t\Omega_0>25$. The torque traces are normalized by $\psiw^2$, and the parameters $\phi$, $\Delta_1$, $l_{21}$ appearing in Eq.~(\ref{eq:Maxwell-torque}) are used as free parameters to align the theoretical curve with the one calculated from the simulation. }
\end{figure}

These observed oscillations are strictly a consequence of the interaction between the destabilized spontaneous mode and the fields driven by the error field, and would be absent without either effect. 
Thus, regardless of how the intrinsic instability emerges, once it forms, it undergoes a `tug-of-war' with the fields driven by the static error field. 
If the error field is not too large, as in the cases under investigation, it only `tickles' the spontaneous mode, allowing it to undergo full orbits in phase space, albeit at a non-uniform rate. 
If the error field is sufficiently strong, it stops the spontaneous mode in its tracks, locking both its phase and the fluid rotation to zero frequency. 
The Appendix presents a quasilinear analytic model that calculates the Maxwell torque on the tearing layer that arises from the interaction of an unstable TM with the fields that are driven by a static error field. 
The resulting torque, given by Eq.~(\ref{eq:Maxwell-torque}), contains the usual contribution from the error field proportional $\psiw^2$, as well as a sinusoidal piece, proportional to the product of the amplitude of the unstable mode with $\psiw$, that oscillates at a frequency of $m\bar{\Omega}_t=2\bar{\Omega}_t$. 
If the spontaneous mode dominates the error field ($|\psit|/\psiw \gg 1$), as is the case in the present NL simulations, the Maxwell torque will be mostly sinusoidal, whereas if the spontaneous mode is absent, the Maxwell torque reduces to the usual steady-state expression proportional to $\psiw^2$. 
The quasilinear torque trace computed from Eq.~(\ref{eq:Maxwell-torque}) is plotted in Fig.~\ref{fig:Torques_vs_time} against the torque calculated from a NL simulation that is driven with $2\psiw=10^{-7}$ and $\Omega_0=2\Omega_r$, the parameters of Figs.~\ref{fig:Psit_width_vs_Omega100_NL} and \ref{fig:Oscillations}).
The horizontal axis of the figure corresponds to the simulation time. This case has $|\psit|\gg |\psiw|$, so that indeed the constant part of the torque is not visible. 
The analytic curve (dashed) is only representative of the Maxwell torque during the post-saturation steady-state phase that begins at $t=4$ for the particular NL simulation used in the comparison. 
As can be seen from the figure, during this phase the two torque traces agree well.
 %------------------------------------------------------------------------
%------------------------------------------------------------------------

The time-asymptotic state that arises out of the NL dynamics described thus far features prominent islands and a large reconnected flux; and yet (average) flow rates that are hardly influenced by the strong NL dynamics, as $\bar{\Omega}_t\simeq \Omega_0$ 
mostly except for the very slowly-rotating cases. 
Thus, we have a time-asymptotic state that appears penetrated in terms of $w$ and $|\psit|$, and somewhat shielded in terms of $\bar{\Omega}_t$.

We performed tests to quantify the effect of the hollow pressure profile outside the tearing layer on the evolution of the driven TM. In principle, this favorable curvature outside the tearing layer can provide stabilization. Specifically, we replaced the hollow (parabolic) pressure profile with a hollow hyperbolic tangent profile, with approximately the same variation in pressure over the tearing layer as the parabolic case. 
Neither the linear nor the NL simulations with the new hyperbolic tangent pressure profile yielded notably different results.

Lastly, the phenomenon described above, which starts with flattening of plasma pressure across the magnetic island, is due to the sound wave, together with pressure advection in the reconnecting flows. 
We expect that thermal conduction would participate in this flattening behavior as well, if it were included in the model. 
The role of anistropic thermal diffusivity in the linear stability of the TM for toroidal geometries as well as the on the stability of the resistive wall mode (RWM) in resistive tokamak plasmas were studied in Refs.~\onlinecite{Connor2015,Xue2018}, respectively. 
Their results show that thermal transport destabilizes the mode--be it the intrinsic TM or a resistive wall mode--by eliminating the stabilizing effect of GGJ. 

%=========================================================================================
\section{Summary and Discussion\label{sec:Discussion}}
%=========================================================================================
This paper investigates the nonlinear behavior of a tearing mode driven by error fields, in a rotating plasma with favorable curvature, by means of resistive MHD simulations. 
The simulations use a large aspect-ratio periodic cylinder with a hollow pressure profile to model a torus with favorable average curvature in the tearing layer. 
The chosen current profile is unstable to the intrinsic $(2,1)$ tearing mode (TM) in the absence of plasma pressure ($\beta=0$), but stabilized for sufficiently large $\beta$.
A nominal flat poloidal rotation profile is imposed with a no-slip boundary condition by means of a momentum source. 
The Lundquist number $S$ is set to $10^5$. 
An error field with helical $(m,n)=(2,1)$ is applied at the boundary to drive the mode. 
The favorable curvature associated with the hollow pressure leads to the propagation of the mode at frequencies $\pm \omega_r$ for a sufficiently large $\beta$ and stabilizes the mode for a larger value of $\beta$ due to favorable average curvature. 
This phenomenon is known as the Glasser or Glasser-Greene-Johnson (GGJ) effect. 

We verify the emergence of the GGJ frequencies and mode stabilization with linear resistive MHD simulations. 
For the parameters in use here, marginal stability occurs at $\beta=0.0014$. 
Linear simulations with $\beta=0.0016$ (weakly stable), a $(2,1)$ error field, and plasma rotation $\Omega_0$ verify that the largest response to the error field (peak reconnected flux) occurs near the phase velocity of the mode $\Omega_r=\pm\omega_r/m$ \cite{Finn2015}. 
At the same time, the maximum response for $\beta=0$ (near $\Omega_0=0$) is strongly screened in the presence of GGJ propagation\cite{Fitzpatrick2017,Liu2012,Finn2015}. 
The present linear simulations also verify that the quasilinear Maxwell torque exerted by the error field on the tearing layer has a qualitatively different dependence \cite{Finn2015} on the plasma rotation frequency from its $\beta=0$ counterpart. 
This leads to the possibility that the plasma might lock to the phase velocity associated with the GGJ effect\cite{Finn2015}. 
The results of these linear simulations with the error field also suggest the existence of the GGJ effect for the visco-resistive (VR) tearing regime. 
%confirm the existence of the GGJ effect for the visco-resistive (VR) tearing regime, since the GGJ effect we observed was seen to occur over a range of parameters for which viscosity and inertia are important.

The major results of the present investigation are all tied to a nonlinear (NL) effect associated with the flattening of the plasma pressure across the magnetic island for small-to-moderate magnitudes of the error field $\psiw$. A major focus is on how this flattening influences the evolution of the weakly stable driven TM. 
The said effect is triggered when $\psiw$ is sufficiently large to cross a threshold in the island width $w=w_c$, above which a secondary instability occurs. 
As a result, the mode grows to a large amplitude, represented by the magnitude of the reconnected flux at the mode rational surface (MRS) $|\psit|$, and dwarfs the error field as well as the peak reconnected flux observed in the linear simulations. 
This secondary instability originates from the destabilization of the spontaneous $(2,1)$ TM: As the error field drives the mode, a magnetic island grows and flattens the pressure within, causing the stabilizing GGJ effect to weaken and eventually vanish. 
The real frequencies of the mode also vanish as a result.  
A similar phenomenon related to spontaneously growing drift-tearing modes was pointed out by Biskamp\cite{Biskamp1979}. 
%This removes the real frequencies of the mode and the stabilization due to the favorable average curvature. 

We demonstrate the destabilization of the spontaneous $(2,1)$ TM by restarting the NL simulations linearly from various times in the parent NL simulation, falling within three critical periods, represented by $w<w_c$, $w_c<w<\delta$, and $w>\delta$, respectively, where $\delta$ is the (linear) tearing layer width. 
Each of these linear simulations uses as its initial equilibrium the symmetric $(m,n)=(0,0)$ fields extracted from various times in the evolution of the parent NL simulations, and advance only an $m=2$ perturbation, representing the mode, with zero error field and no plasma rotation. 
The linear restarts corresponding to the period when $w_c<w<\delta$ show no GGJ oscillations and a linear instability with a growth rate that is comparable to the instantaneous growth rate of the mode calculated from the logarithmic derivative of $|\psit(t)|$ in the parent simulation. 
The linear restarts corresponding to periods when $w<w_c$ and $w>\delta$ show stability, consistent with the observed evolution of the mode in the parent NL simulation. 
%Cases which saturate linearly with $w<w_c$ show GGJ oscillations and weak damping. 
%For cases that correspond to $w_c<w<\delta$ in the parent NL simulation, a linear instability without GGJ oscillations is observed. 
%The growth rate extracted from these linear simulations match the instantaneous growth rate of the mode calculated from the logarithmic derivative of $|\psit(t)|$
%During this phase, the pressure is observed to flatten across the tearing layer, which destabilizes the spontaneous $(2,1)$ tearing mode. 
%The linear restarts corresponding to the Rutherford phase, i.e. $w\approx \delta$ in the parent NL simulation show weakened growth, in conjunction with the moderate flattening of the current density $J_z^{(0,0)}$ observed in the parent simulation. 
%Finally, the linear restarts show complete stabilization for the fields extracted from the nonlinear saturation phase of the parent simulation, associated with flattening of the equilibrium current across the magnetic island ($w>\delta$). 
Since the effects of the pressure in the layer disappear when $w\approx w_c$, the evolution of the mode from this point on proceeds in a manner similar to that of the $\beta=0$ case with the same $q(r)$ profile.   
For this reason, this weakly-driven and weakly-damped finite-$\beta$ TM grows to the same amplitude as a $\beta=0$ unstable TM that is driven by the same error field. 

A consequence of the above secondary instability is the emergence of oscillations, starting during the nonlinear saturation phase, at twice (because $m=2$) the average fluid rotation frequency at the MRS. These oscillations appear in the phase velocity of the mode, in the plasma rotation rate throughout the volume, in the island width, and in the mode amplitude at the MRS. 
%A consequence of the above phenomenon is the appearance of oscillations, at the initial rotation frequency $\Omega_0$, which start as the secondary instability develops large amplitude and develop fully after the nonlinear saturation phase. In this later phase, oscillations are observed in the phase velocity $\Omega_{ph}$ of the mode and the fluid speed throughout the volume and specifically $\Omega(r_t)$. At the same time, there are oscillations in the island width $w(t)$, and mode amplitude $|\psit(t)|$. 
These oscillations are strictly a consequence of the interaction between the destabilized spontaneous mode and the static fields driven by the error field, and require the presence of both effects. 
The quasilinear analytic model presented in the Appendix illustrates how an oscillating Maxwell torque can arise out of the interaction of an unstable TM with static fields driven by an error field.  
Our findings suggest the exciting possibility that these oscillations might be observable in toroidal confinement experiments, featuring RMP's or error fields. 
The oscillations observed here are not the same as the pulsations described in Ref.~\onlinecite{Fitzy2018}, as the island width here has a small oscillation amplitude and certainly never shrinks back to zero width; also, the phase of the mode undergoes full orbits in the complex plane. Furthermore, the pulsations observed in Ref.~\onlinecite{Fitzy2018} are two-fluid in nature, whereas the present oscillations occur in resistive MHD and do not require a pressure gradient across the island region.

The range in the magnitude of the error field $\psiw$ where linear behavior gives way to this nonlinear oscillatory behavior is limited to roughly an order of magnitude for the present parameters. 
Above this range in $\psiw$, the aforementioned oscillations disappear and the conventional locking of the phase velocity of the mode and the fluid rotation at the MRS to zero frequency occurs.

The values of the critical width $w_c$ signaling destabilization of the spontaneous secondary mode are determined empirically here based on the results of the linear restarts. 
These values are approximately three times the value based on a GGJ analog of the criterion applied to frequencies $\omega_r \sim \omega_*$ for drift-tearing modes in Ref.~\onlinecite{Scott1985}. 
The critical width $w_c$ is also a fraction of the linear layer width $\delta$ for the present parameters, implying that the pressure flattening precedes NL saturation by current flattening for this intermediate range of Lundquist number $S$. 
However, since $\delta$ scales inversely as a fractional power of $S$, at least in the resistive MHD tearing regimes, it is possible to reverse the ordering such that $w_c>\delta$ for $S\gtrsim 10^7$. 
In that case the mode should begin to saturate due to current-flattening before the island grows large enough to turn off the stabilizing GGJ effect. 
This reversal for larger $S$ also suggests the possibility of plasma locking to a finite frequency\cite{Finn2015}, in this case the real frequency of the mode, instead of the conventional locking to zero frequency. 
This phenomenon of finite-frequency locking is currently under study and will be the topic of a future publication. 
%\textcolor{blue}{The motivation here is to compare with the analog of the result in Ref.~\onlinecite{Scott1985}, where the critical island width is defined with $c_s$ and $\omega_*$; the resistive MHD analog has $\omega_* \rightarrow \omega_r$, the latter due to the Glasser effect. The said work indicates that when the magnetic island width exceeds this critical width, the real frequency of the mode associated with $\omega_*$ slows down and eventually ceases as the expanding island flattens the plasma pressure. However we found that this analytic critical width underestimates the observed $w_c$ by a factor of three. }

This is a dynamic problem that requires self-consistent nonlinear simulations to understand the interplay between the nonlinear effects of pressure-flattening, current-flattening, and locking. 
It cannot be captured by linear simulations or ad-hoc models of pressure-flattening. 
Furthermore, at a very large value of $S$ an accurate description of the tearing layer physics might require two-fluid, finite-Larmor, and even some kinetic effects because of the very small layer width. 
%It is evident that our model does not address the poloidal mode coupling %akin to a toroidal geometry, which might matter in the presence of %multiple tearing layers. 
A more comprehensive model must also involve toroidal geometry with a peaked pressure profile. In such a model, additional related toroidal effects such as pressure flattening across multiple rational surfaces and stochastic field line regions, especially near the edge, should occur.
An investigation combining all these elements is the natural extension of the present work as well as past works published in this area and may be crucial in improving our understanding of how a rotating tokamak plasma interacts with non-axisymmetric external magnetic fields.

%=============================================================================
\section*{Acknowledgements}
%=============================================================================
The work of C.~Ak\c{c}ay, J.~M.~Finn, A.~J.~Cole, and D.~P.~Brennan was supported by the DOE Office of Science collaborative
Grant Nos. DE-SC0019016, DE-SC0014119, and DE-SC0014005, respectively. 
We thank Carl Sovinec at the University of Wisconsin Madison for his help with the NIMROD code. 
This research used resources of the National Energy Research Scientific Computing Center (NERSC), a U.S. Department of Energy Office of Science User Facility operated under Contract No. DE-AC02-05CH11231.

%=============================================================================
\appendix
\section*{Appendix:Torque on a tearing layer in the presence of plasma rotation, an
error field and an unstable spontaneous mode}
%=============================================================================
In this appendix we present an analytic calculation of the torque on a tearing layer in the presence of an error field, plasma rotation, and a spontaneously growing tearing mode (TM). 
We are not concerned here with how the unstable TM came to be -- in this paper, the mode becomes destabilized due to the flattening of the pressure profile across the magnetic island. Here, we simply assume that it is present. 
%For application to the present paper, the spontaneous mode is unstable due to flattening of the pressure profile. 
%As discussed in Sec.~\ref{sec:NLRuns}, this flattening removes the real frequency and stabilization due to the Glasser effect.
For simplicity, we assume that the tearing layer is in the constant-$\psi$ VR regime, appropriate if the pressure is flattened and the mode grows slowly.

The Maxwell force density in reduced MHD is given by $j_{z}B_{r}$ and in the layer at the mode rational surface (MRS) is proportional to $M=\partial_{x}q_{m}(x)=q_m'(x)$, where\cite{Cole2015}
\[
q_{m}=-\text{Im}\left(\tpsi^{*}\tpsi'\right);
\]
here $\tpsi$ represents the perturbed or reconnected flux. 
The net torque on the layer at $r=r_{t}$ is proportional to
\begin{equation}
N_{m}=\int_{r_{t}-\delta/2}^{r_{t}+\delta/2}Mdx=[q_{m}]=-\text{Im}\left(\psit^*[\tpsi']_{r_t}\right),\label{eq:Force}
\end{equation}
where $x=r-r_t$, $\delta$ is the full width of the layer and $[\cdot]_{r_{t}}$
is the jump across the layer. We assume a spontaneous TM, purely growing
in the plasma frame and with amplitude $\alpha_{1}(t)$, and an error
field response driven by the error field $\psiw$, with amplitude
$\alpha_{2}(t)$. We assume that the second set of fields do not reconnect
at $r_{t}$ on the growth time scale. The fields are represented as
\[
\tpsi=\alpha_{1}(t)\phi_{1}(r)+\alpha_{2}(t)\phi_{2}(r),
\]
where $\phi_{1}(r_{t})=1,\,\phi_{1}(r=a)=0,\,\,[\phi_{1}']_{|r_{t}}=\Delta_{1}$
and $\phi_{2}(r_{t})=0,\,\,\phi_{2}(a)=1,\,\,[\phi']_{|r_{t}}=l_{21}\doteq\phi_{2}'(r_{t}+)$.
For the VR regime with time constant $\tau$, we have the conditions
$\tau\left(\partial/\partial t+i\Omega\right)\tpsi(r_{t})=[\tpsi']_{|r_{t}}$and
$\psiw=\tpsi(a)=\alpha_{2}$, where $\Omega=m\bar{\Omega}_t=2\bar{\Omega}_t$ is the Doppler shift frequency at $r=r_t$, based on a time-average of the fluctuating flow $\Omega_t$ there. (In this calculation we neglect the quasilinear effect of the Maxwell force on the rotation, so we can assume $\bar{\Omega}_t=\Omega_0$).

We find, with $\tau=1$,
\begin{equation}
\left(\frac{\partial}{\partial t}+i\Omega-\Delta_{1}\right)\alpha_{1}=l_{21}\psiw.\label{eq:Simplest-equation-for-alpha1}
\end{equation}
 Integrating over $t$, this yields $\tpsi(r_{t},t)=\alpha_{1}(t)$, where
 \begin{align}
\alpha_{1}(t)&=\hspace{-1mm}\left(\alpha_{1}(0)-\frac{l_{21}\psiw}{i\Omega-\Delta_{1}}\right)e^{(\Delta_{1}-i\Omega)t}+\frac{l_{21}\psiw}{i\Omega-\Delta_{1}}\nonumber,\\
&=\hat{\psi}_{t}e^{-i\Omega t}+\frac{l_{21}\psiw}{i\Omega-\Delta_{1}},\label{eq:recon-flux}
\end{align}
where the unstable mode is represented by $\hat{\psi}_{t}$, which includes the growth factor $e^{\Delta_{1}t}$.
If $\Delta_{1}<0$ the solutions spirals in to the steady-state solution
$\alpha_{1}(t)=l_{21}\psi_{w}/(i\Omega-\Delta_{1})$. 
This is the steady-state amplitude of the mode at the MRS for the usual driven-reconnection problem. 
If $\Delta_{1}>0$ the steady-state solution is present and the spiraling part grows.

For the second factor in Eq.~(\ref{eq:Force}) we have 
\begin{equation}
[\tpsi']_{\vert r_{t}}=\Delta_{1}\hat{\psi}_{t}e^{-i\Omega t}+\frac{i\Omega l_{21}\psiw}{i\Omega-\Delta_{1}}.\label{eq:Jump-tearing-layer}
\end{equation}
We find $-\tpsi^{*}(r_{t})[\tpsi']_{\vert r_{t}}$ equals
\begin{align}
\hspace{-9mm}&-\left(\hat{\psi}_{t}^{*}e^{i\Omega t}-\frac{l_{21}\psiw}{i\Omega+\Delta_{1}}\right)\hspace{-2mm}\left(\Delta_{1}\hat{\psi}_{t}e^{-i\Omega t}+\frac{i\Omega l_{21}\psiw}{i\Omega-\Delta_{1}}\right)\nonumber\\
&=-\Delta_{1}|\hat{\psi}_{t}|^{2}-l_{21}\psiw\left(\frac{i\Omega\hat{\psi}_{t}^{*}e^{i\Omega t}}{i\Omega-\Delta_{1}}-\frac{\Delta_{1}\hat{\psi}_{t}e^{-i\Omega t}}{i\Omega+\Delta_{1}}\right)\nonumber\\
&-\left(\frac{i\Omega l_{21}^{2}\psi_{w}^{2}}{\Omega^{2}+\Delta_{1}^{2}}\right).
\end{align}
Writing $\hat{\psi}_{t}=|\hat{\psi}_{t}|e^{i\phi}$ and
using Eq.~(\ref{eq:Force}), we obtain
\begin{equation}
N_{m}=-\frac{ml_{21}\psi_{w}|\hat{\psi}_{t}|}{2}\sin(\Omega t-\phi)-\frac{m}{2}\left(\frac{\Omega l_{21}^{2}\psi_{w}^{2}}{\Omega^{2}+\Delta_{1}^{2}}\right).\label{eq:Maxwell-torque}
\end{equation}
The first term $\propto|\hat{\psi}_{t}|^{2}$, strictly due to the unstable (rotating) tearing mode and interacting only with itself, has disappeared. 
The error field term $\propto\psi_{w}^{2}$, has the usual VR (Lorentzian) form and is due to the phase shift relative
to the applied error field because of the plasma rotation $\Omega$. 
The coupling term ($\propto\psi_{w}|\hat{\psi}_{t}|$) is sinusoidal with a frequency of $\Omega$. 
This is exactly the term that is responsible for the oscillations described in Section~\ref{sec:NLRuns}, with the substitution $\Omega\rightarrow m\bar{\Omega}_t=2\bar{\Omega}_t$.  

The relative strength of these two terms depends on the ratio of $\hat{\psi}_{t}$
to $\psi_{w}$ as well as on $\Omega$ and $\Delta_{1}$. For a faster
growing tearing mode, the $\psi_{w}^{2}$ term becomes \textit{less}
peaked near $\Omega=\Delta_{1}$. If the $\psi_{w}|\hat{\psi}_{t}|$
term dominates, the plasma rotation at the MRS and the phase velocity
of the perturbed fields is modulated by the oscillating torque. These
effects, in which the torque affects the rotation rate, are outside
the scope of this calculation and are discussed in Section~\ref{sec:NLRuns}.


\begin{thebibliography}{99}
\bibitem{FootNote5}
The {NIMROD MHD} code does not have the capability to advance a single helicity
  pair $(m,n)$ because it uses 2{D} (r-z) finite elements. {T}he $n=1$
  amplitude is calculated a posteriori via an {FFT} of the $m=2$ mode.
\bibitem{FootNote1}
In cylindrical geometry, assuming zero parallel velocity $\vpar=0$, we find
  $v_{\theta}^{(cyl)}=-{E}_{r}{B}_{z}/{B}^{2}$ with
  $v_{z}^{(cyl)}/v_{\theta}^{(cyl)}=\mathcal{O}(\epsilon)$ where
  $\epsilon=a/{R}$. On the other hand, modeling strong poloidal flow damping in
  a torus, we take $v_{\theta}^{(tor)}=0$ rather than $\vpar=0$, giving
  $v_{z}^{(tor)}={E}_{r}/{B}_{\theta}$. taking
  $\capom_{\theta}=v_{\theta}^{(cyl)}/r$ and $\capom_{\phi}=v_{z}^{(tor)}/{R}$
  we find
  $\capom_{\theta}/\capom_{\phi}=-{B}_{\theta}{B}_{z}{R}/{B}^{2}r=\mathcal{O}(1)$.
  in other words, the doppler shift due to poloidal rotation in cylindrical
  geometry is nearly the same as the doppler shift due to toroidal rotation in
  toroidal geometry.
\bibitem{FootNote6}
We use only the equilibrium component of the plasma density in the momentum
  equation, to keep only quadratic coupling of the fourier modes. this does not
  change the results of the nonlinear simulations qualitatively.
\bibitem{FootNote2}
A damped $\beta=0$ {RI} mode also admits two maxima in the reconnected flux,
  which are however much closer together than the two peaks of the finite
  $\beta$ {TM} reconnected flux curve. {C}onsequently, the doubly-humped
  response of the {RI} tearing mode also causes screening at zero frequency,
  but only weakly.
\end{thebibliography}
\end{document}